\documentclass[12pt]{article} 

\usepackage[parfill]{parskip}    
\usepackage{graphicx}
\usepackage{epstopdf}
\usepackage{amsmath, amssymb, amscd}

\begin{document}

\title{Is it possible to control the spread of a globalized culture?}
\author{Nino Boccara\\
Department of Physics, University of Illinois at Chicago\\
and\\
DRECAM/SPEC, CE Saclay, 91191 Gif-sur-Yvette Cedex, France\\
\texttt{boccara@uic.edu}}

\date{}

\maketitle

\begin{abstract}
We study a model describing the spread of a globalized culture in a population of individuals localized at the nodes of a social network. The influence of this globalized culture, assumed to be foreign to the local culture, is measured by a probability to convince each individual to adopt its cultural traits. This probability depends upon the degree $s$---a real between $0$ and $1$---of ``wise'' skepticism characterizing the personality of each individual and a parameter $a$ representing the resistance of the society as a whole to the spread of the foreign cultural traits.  A greater $a$ indicates a stronger resistance of the local culture to globalization. On the other hand, each individual interacts with a random number of other individuals---her cultural neighborhood---uniformly distributed between 1 and a maximum value. The probability distribution of an individual to belong to the cultural neighborhood of another individual has a power-law behavior. A small fraction $r$ of the total population belonging to the tail of this probability distribution have an $s$-value equal to $1$. They represent the most conservative individuals firmly attached to their local culture.
\end{abstract}

\section{Introduction}

Globalization refers to the growing integration of economies and societies over the past two decades~\cite{lb2003}. This process, made possible by the rapid advances in communication technologies, has created a smaller world in which ideas and money can move across borders almost instantly.
 
Since the early 1990s, globalization has triggered a storm of controversy among social activists, intellectuals, business leaders, policy makers and politicians. The debate over globalization is often passionate and, sometimes, even violent. Proponents~\cite{tf2000, imf2000, jb2004, mw2004} consider market liberalism as highly productive, raising living standards worldwide, and  favoring innovations, but critics~\cite{wg1997, jmeg2001, rs2003} argue that globalization creates inequalities, loss of jobs, environmental degradation, and ``is a declaration of war upon all cultures.''~\cite{js2004} 

Globalization has a great impact on culture. It destroys local traditions in order to create a homogenized world culture.  Culture is a multifaceted concept characterizing the behavior acquired through social interactions of human beings as members of a society. It involves, at least, three components: what these individuals think, what they do, and the material products they produce~\cite{nasp}. Many sociologists have discussed the influence of globalization on culture.

According to Samuel Huntington~\cite{sh1993, sh1996}, the  ``great divisions among humankind and the dominating source of conflict will be cultural.''  Though many scholars agreed with Huntington, many also criticized his aggressive picture of the non-Western civilizations while ignoring the misdeeds of the Western civilization~\cite{tc1993}. These misdeeds of the Western civilization have been depicted by George Ritzer as \emph{The McDonaldization of Society}~\cite{gr1993, gr2002, gr2004}. For Ritzer, McDonaldization is rationalization taken to an extreme level with the consequence that creativity is taken out of all activities, turning them into a series of routinized kinds of procedures. For Benjamin Barber~\cite{bb1995}, the alternative of the aggressive economic and cultural globalism of McWorld is Jihad, understood not as Islam but as disintegral tribalism and reactionary fundamentalism.  While McWorld eliminates our differences, Jihad overemphasizes those differences. The future of international relations is dominated by their conflict. 

When you visit a mall, sit in a multiplex movie theater, sleep in a modern hotel or eat in a fast-food restaurant, it is not easy to figure out where you are. McWorld requires uniformity to maximize profits. 
But, is it possible to protect cultural diversity, which ``is as necessary for humankind as biodiversity is for nature,''  according UNESCO Universal Declaration on Cultural Diversity?\footnote{This declaration has been adopted in Paris on November 2, 2001. It can be found at: http://europa.eu.int/comm/avpolicy/extern/gats2000/decl-en.pdf.}  

In September 1993 Europeans, led by the French,  demanded that trade in audiovisual products be left outside the GATT agreement, arguing that cultural products cannot be treated as ordinary commercial products. The idea behind this so called \emph{exception culturelle} (cultural exception) was that without imposing restrictions to control the flow of cheap American products onto European markets, European culture would be threatened. European protectionism, of course strongly resisted by the US, was seen as an essential condition to prevent all the globe from becoming Disneyland.

Between cultural differentialism or lasting difference  and cultural convergence or growing sameness, there is, according to Nederveen Pieterse~\cite{jnp2004}, a third way: cultural hybridization, global m\'elange, or ongoing mixing.  According to Benjamin Barber, apparent hybridization is just be thinly disguised globalized monoculture~\cite{bb1995}:  ``McDonald's \emph{adapts} to foreign climes with wine in France even as it imposes a way of life that makes domestic wines irrelevant.''

\section{Model's Description}

In order to understand if cultural diversity can be protected, we have studied a model for the spread of a globalized culture in a population of individuals located at the vertices of a scale-free social network. In this paper the social network is a directed graph, that is, an ordered pair of disjoint sets $(V,E)$, where $V$ is a nonempty set of elements called vertices, nodes, or points, and $E$ a set of ordered pairs of distinct elements of $V$, called  directed edges, arcs, or links.  Most social networks~\cite{bjnrsv2002, r1998, leasa2001, emb2002,ab2002} are characterized by a small average shortest path length between two randomly selected vertices, a high clustering coefficient and a power-law probability distribution for the vertex degrees.

Following Axelrod~\cite{ra1997} an individual's culture is represented by an ordered list of features such as dress style or customary food. Each feature has a set of traits, which are the alternative values the feature may have.  If the culture is defined by $f$ different features, each taking $n$ different values, the culture of every individual is described by a sequence of $f$ integers ranging from $1$ to $n$. In our model, the globalized culture is assumed to be foreign to the local culture and has, therefore, cultural trait values not shared by the population. Conventionally we characterize the globalized culture by a sequence of $f$ zeros.  

Each individual's personality is represented by a real number $s$ between 0 and 1, which may also be viewed as her social status. An individual with a high $s$-value has a strong convincing power when interacting with other individuals and a high degree of ``wise'' skepticism when interacting with the foreign culture. 

In order to mimic cultural exception, that is, the resistance to the spread of the foreign culture,  the interaction of the individuals with the foreign culture is characterized by a positive exponent $a$.  A larger $a$ indicates a stronger resistance of the local culture to globalization. The exponent $a$ can also be viewed as a measure of the effort made by the society to teach and promote its specific culture.

More precisely, our social network model is represented by a directed graph with $N$ nodes. Each node is occupied by an individual characterized by
\begin{itemize}
\item[1-] her \emph{name}: an integer between $1$ and $N$,
\item[2-] her \emph{$s$-value}: a real between $0$ and $1$,
\item[3-] the list of her \emph{cultural neighbors}: a list of other individuals with whom she interacts,\footnote{The cultural neighbors of an individual are not necessarily people that person physically meets, they could be, for instance, writers or movie directors attached to local cultural traits values.} 
\item[4-] her \emph{culture}: a list of $f$ trait values, ranging from $1$ to $n$.
\end{itemize} 
The size of each neighborhood is a random integer uniformly distributed between $1$ and the maximum cultural neighborhood size, here taken equal to either $8$ or $17$, An individual $k$ belongs to the cultural neighborhood of individual $i\neq k$ if the directed link $(i,k)$ belongs to the set $E$ of graph's edges.  While vertex out-degrees are uniformly distributed between $1$ and the maximum cultural neighborhood size, the random selection of neighbors is such that vertex in-degrees have a Pareto probability distribution with a minimum value parameter equal to $1$ and a shape parameter equal to $2$,\footnote{The cumulative distribution function of the Pareto distribution with a minimum value parameter equal to $x_0$ and a shape parameter equal to $\sigma$ is
$$
x\mapsto 1-\left(\frac{x_0}{x}\right)^\sigma.
$$}
that is,  the Pareto cumulative distribution function $x\mapsto 1-x^{-2}$, and, consequently, a probability density function $x\mapsto 2x^{-3}$. We shall call such a social network, a \emph{Pareto society}. A fraction $r$ of the total number $N$ of individuals, selected among those having the highest in-degrees, defines the subnetwork of \emph{cultural leaders},  that is, the subnetwork of the most influential individuals whose $s$-value is set equal to $1$. For example, in an initial configuration with a fraction of cultural leaders $r=0.1$, a maximum cultural neighborhood size equal to $8$ and a culture defined by only two features ($f=2$), each taking only two possible different trait values ($1$ or $2$), the state of individual $789$ is
$$
\big(789, 0.094619, (1396, 579, 2117, 2143, 148, 702), (2, 1)\big).
$$ 
She has an $s$-value equal to $0.094619$, her cultural neighborhood, of size $6$,  is 
$$
(1396, 579, 2117, 2143, 148, 702),
$$
 and her culture is represented by the pair of trait values $(2,1)$. 
 
The global evolution rule of the social network consists of two synchronous subrules defined as follows.
\begin{itemize}
\item[1-] \emph{Imitate rule}: At each time step, every individual locates the person in her neighborhood who has the highest $s$-value and changes the value of one of her randomly selected feature trait to equal the value of the corresponding trait value of that person if, and only if, her $s$-value is less than the $s$-value of that person. 
\item[2-] \emph{Globalizing rule}: At each time step, a randomly selected feature trait of each individual is changed to $0$ with a probabilty $(1-s)^a$, where $s$ is the $s$-value of the individual and $a$ the exponent characterizing the resistance of the society to the spread of the globalized foreign culture.
\end{itemize}
If it were not for the imitate rule, under the action of the globalizing rule, all individuals having an $s$-value less than $1$, that is, the majority, will eventually adopt all the trait values of the foreign culture. The globalizing rule mimics the role of the huge advertising power of multinational corporations conquering a new national market. Only the fraction $r$ of the most influential individuals, with an $s$-value equal to $1$, will, in any case, never adhere to the foreign culture. They represent a small fraction of conservative individuals firmly attached to their culture keeping unchanged their cultural traits  under the action of the evolution rule. 

\begin{figure}[h]
\begin{center}
\scalebox{0.7}{\includegraphics{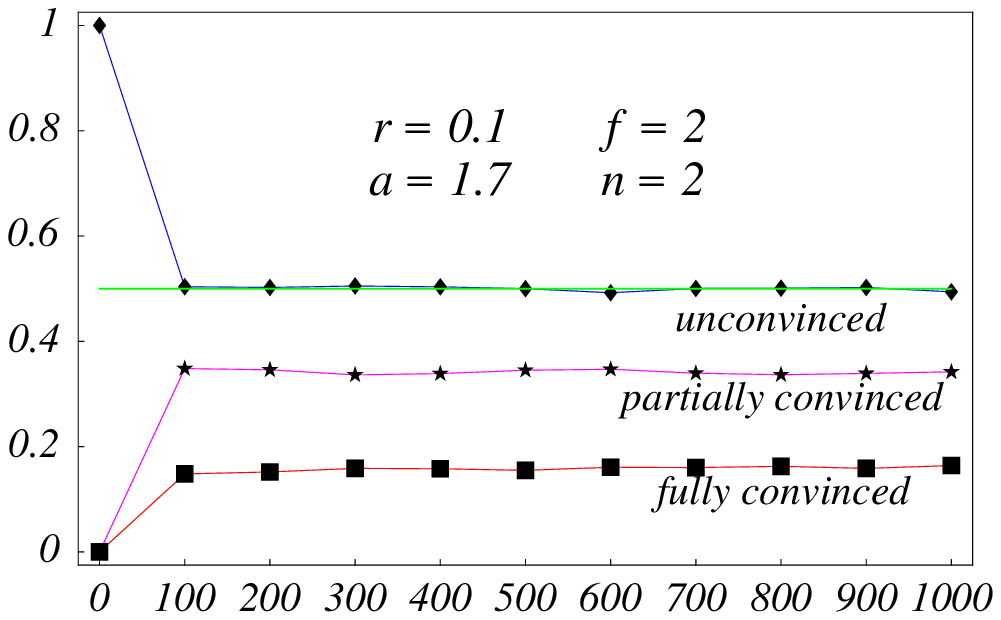}}
\end{center}
\vspace{0.2cm}
\begin{center}
\scalebox{0.7}{\includegraphics{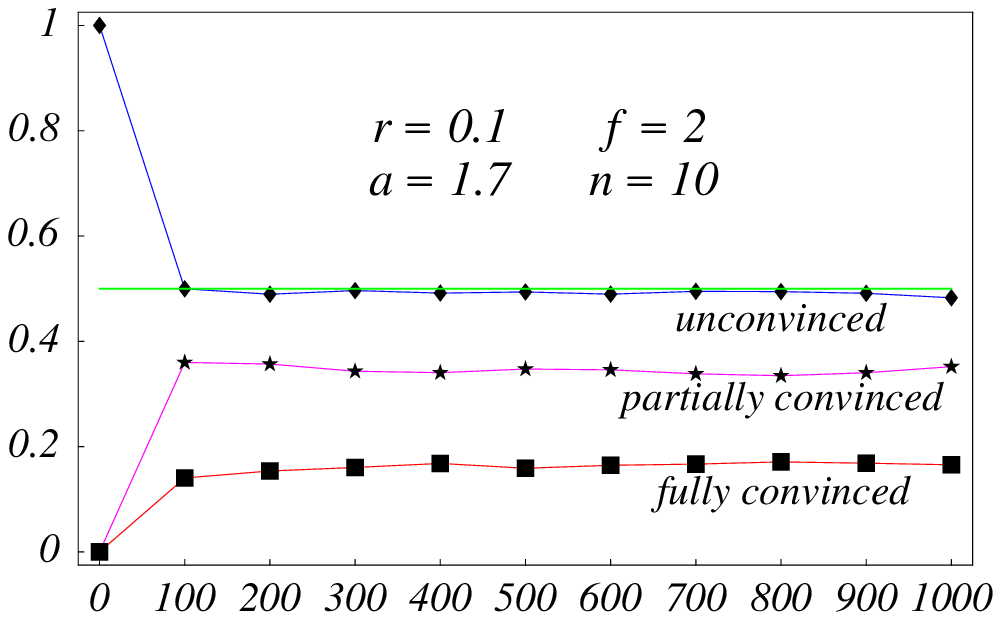}}
\end{center}
\caption{ Fraction of the total number of individuals adhering to the foreign culture in the case of a society with a fraction of cultural leaders $r=0.1$, and a culture characterized by only $2$ features with either $2$ (top) or $10$ (bottom) traits \emph{per} feature.}
\label{fig:a17rho10f2}
\end{figure}

Note that this model can be adapted to study related problems such as, for example, the struggle between local brands of a given product (drink, cosmetic, appliance) and a new brand trying to conquer the local market. 

\section{Numerical Results}

In all our simulations the social network consists of $2500$ individuals.  The steady state is reached rather rapidly, after about a few hundreds time steps.

Consider first the case of a society with a fraction of cultural leaders $r=0.1$, a cultural neighborhood maximum size equal to $8$, and a culture characterized by only $2$ features. Whatever the number of different traits \emph{per} feature, the numbers of fully convinced, partially convinced and unconvinced individuals are not modified. This is not surprising since, in this model, the probability for an individual to adopt the trait value of the foreign culture does not depend upon her original trait value. As shown in Figure~\ref{fig:a17rho10f2}, for $a=1.7$, approximately half of the total population either partially or fully adheres to the foreign culture. 

\begin{figure}[h]
\begin{center}
\scalebox{0.7}{\includegraphics{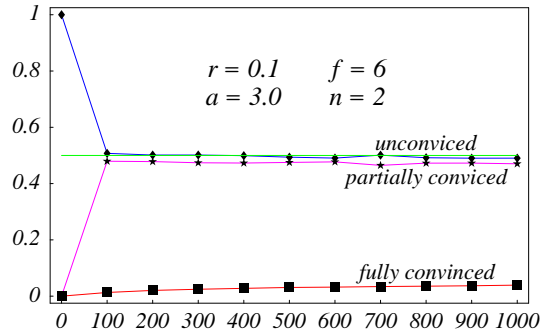}}
\end{center}
\caption{Fraction of the total number of individuals adhering to the foreign culture in the case of a society with a fraction of cultural leaders $r=0.1$, an exponent $a=3.0$, and a culture characterized by $6$ features and $2$ traits \emph{per} feature.}
\label{fig:rho10a30f6t2}
\end{figure}

If we increase the number $f$ of features, we have to increase the resistance of the local culture to the foreign culture, that is, increase the exponent $a$, to keep the same number of individuals rejecting the foreign culture (see Figure~\ref{fig:rho10a30f6t2}). When the number of features increases, the number of fully convinced individuals, that is, the number of individuals adopting \emph{all} the trait values of the foreign culture, sharply decreases since the probability for an individual to adopt all the feature trait values of the foreign culture decreases exponentially with the number of features.

\begin{figure}[h]
\begin{center}
\scalebox{0.7}{\includegraphics{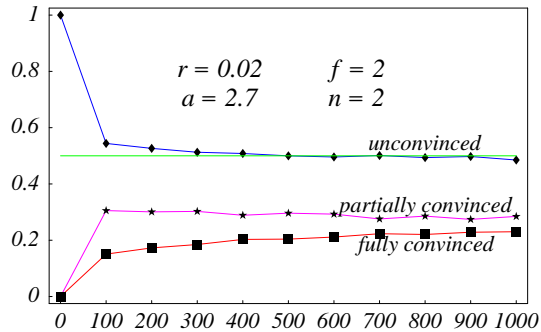}}
\end{center}
\caption{Fraction of the total number of individuals adhering to the foreign culture for $a=2.7$ in the case of a society with a fraction of cultural leaders $r=0.02$, and a culture characterized by only $2$ features and $2$ traits \emph{per} feature.}
\label{fig:rho02a27f2t2}
\end{figure}

If we decrease the fraction $r$ of cultural leaders, clearly we have to increase the exponent $a$ for the foreign culture not to be able to convince more than half of the population accepting the new culture traits. For $a=2.7$, as shown in Figure~\ref{fig:rho02a27f2t2}, approximately half of the total population either partially or fully adheres to the foreign culture. The fraction  $r$ of cultural leaders being equal to $0.02$,  in this case, $48$~\% of the total number of indiduals with an $s$-value less than 1 totally rejects the foreign culture.

\begin{figure}[h]
\begin{center}
\scalebox{0.7}{\includegraphics{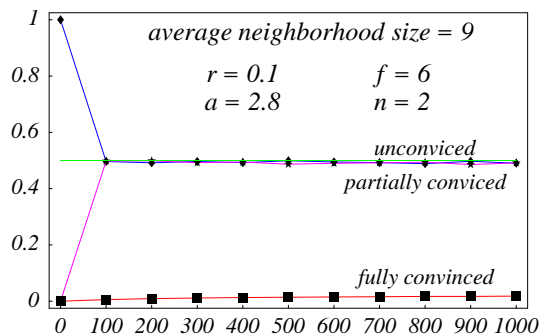}}
\end{center}
\caption{Fraction of the total number of individuals adhering to the foreign culture for $a=2.8$ in the case of a society with a fraction of cultural leaders $r=0.1$, a randomly uniformly distributed neighborhood size between $1$ and $17$, and a culture characterized by $6$ features and $2$ traits \emph{per} feature.}
\label{fig:rho10a28ms17f6t2}
\end{figure}

All these results were obtained with a cultural neighborhood maximum size equal to $8$, that is, an average number of cultural neighbors equal to $4.5$. If we increase this number, the individuals become more connected and, consequently, are able to resist more effectively to the influence of the foreign culture. Doubling the average number of neighbors, that is, choosing a cultural neighborhood maximum size equal to $17$,  for $r=0.1$, $f=2$, and $n=2$, the value of exponent $a$ necessary to keep the sum of the numbers of fully convinced and partially convinced individuals not larger than half of the total population decreases slightly from $3.0$ to $2.8$ as shown in 
Figure~\ref{fig:rho10a28ms17f6t2}.

\section{Results Analysis}

What does differentiate the different groups of individuals? In the case of a society with a fraction of cultural leaders $r=0.1$, 6 features with 2 traits each, and an average cultural neighborhood size equal to 4.5. For $a=3.0$ we find approximately that 
\begin{itemize}
\item $2.6$ \%  of the population fully adopt the foreign culture,
\item  $47.4$ \% of the population partially adopt the foreign culture,
\item  $40$ \% of individuals with an $s$-value less than $1$ completely reject the foreign culture, and
\item $10$ \% leaders who, in any case, never adopt any foreign cultural trait. 
\end{itemize}

Individuals fully adopting the foreign culture are characterized by a small number of neighbors, $1.4$ \emph{per} individual on average, and among these neighbors no cultural leaders are present. That is, individuals easily influenced by the foreign culture are weakly connected to other individuals, and have no interactions with cultural leaders. 

Individuals partially adopting the foreign culture have an average of $2.67$ out of $6$ traits equal to $0$. Their average neighborhood size, which is equal to $4.4$, includes an average number of cultural leaders equal to $1.47$, allowing them to make up for their small $s$-values---which can even be slightly smaller than the $s$-values of individuals fully adopting the foreign culture (in our simulations, $0.25$ compared to $0.33$ on average). 

In a society with a culture defined by $6$ different features,  the individuals adopting all the culture traits of the foreign culture are not characterized by the lowest $s$-value. They may even have a slightly higher $s$-value than the individuals who only partially adopt the foreign cultural traits. Their essential characteristic is the small size of their cultural neighborhood which, moreover, does not include any cultural leader.

Individuals completely rejecting the foreign culture are characterized by a high $s$-value---0.88 on average---and have, on average, a number of cultural neighbors equal to $4.67$ including $1.68$ cultural leaders.

\section{Conclusion} We have studied a model describing the spread of a globalized culture in a population of individuals localized at the nodes of a social network. At each time step, under the influence of the cultural neighbors a person interacts with, this person, whose social status is measured by a parameter $s$ between $0$ and $1$, may modify one of her cultural traits to adopt the corresponding trait of her neighbor having the greatest $s$-value. A small fraction $r$ of the total population are cultural leaders, champions of their local culture who have a very strong convincing power when interacting with other individuals. Moreover, all individuals are under the influence of a foreign culture whose probability to convince a particular individual is a decreasing function of her $s$-value and the exponent $a$ characterizing the resistance of the society as a whole.  As a result of both interactions, the individuals may be divided into three groups. The \emph{fully convinced} group consists of individuals who, in the steady state, have fully adopted all the cultural traits of the foreign culture. They are essentially individuals who interact with very few other individuals and, in particular, do not interact with the fraction $r$ of cultural leaders, promoting champions of the local culture.  The \emph{partially convinced} group are individuals who interact with a number of other individuals close to the average value of neighborhood sizes with among them a reasonable number cultural leaders, but their average $s$-value is small, making them easy to influence. The individuals of the \emph{unconvinced} group have the same characteristics as the individuals of the previous group except that they have a very high average $s$-value.  

In essence, this model shows that to resist globalization and keep alive a diverse local culture, role models are essential, creations of as many links as possible between individuals sharing the same local culture has to be encouraged, and the society as a whole must make every effort to protect and promote its own culture even at the cost of not fully respecting the rules of free trade.

\end{document}